# Atomic-scale analysis of disorder by similarity learning from tunneling spectroscopy

Petro Maksymovych[1], Jiaqiang Yan[2], Brian Sales[2], Jun Wang[1]

[1]Center for Nanophase Materials Sciences, Oak Ridge National Laboratory, Oak Ridge, TN, 37831
[2]Materials Science and Technology Division, Oak Ridge National Laboratory, Oak Ridge, TN, 37831

**Abstract**

Rapid proliferation of hyperspectral imaging in scanning probe microscopies creates unique opportunities to systematically capture and categorize higher dimensional datasets, toward new insights into electronic, mechanical and chemical properties of materials with nano- and atomic-scale resolution. One of the central goals of hyperspectral imaging is the development of a consistent framework for data analysis that would be effective, reproducible and transferrable in resolving the structure of the datasets, conceptually resembling the success of integral transforms in image analysis. Here we demonstrate a similarity learning for tunneling spectroscopy acquired on superconducting material (FeSe) with sparse density of imperfections (Fe vacancies). Popular methods for unsupervised learning and discrete representation of the data in terms of clusters of characteristic behaviors were found to produce inconsistencies with respect to capturing the location and tunneling characteristics of the vacancy sites. The underlying reason for their ambiguity was traced to continuous variation of the electronic properties across the surface and therefore absence of clear structural boundaries in the low-dimensional latent spaces of the data. We supported this hypothesis by direct analysis of the distributions of Euclidean distances within the dataset, and proposed distance rescaling with probabilistic description as a possible approach to mitigate the detrimental effect of the long tails of the distributions on the performance of clustering methods. Subsequently, we applied a more general, non-linear similarity learning, where dimension reduction was explicitly trained to amplify similarities and dissimilarities between data-points. This approach was found to outperform several widely used methods for dimensionality reduction and produce a clear differentiation of the type of tunneling spectra. In particular, significant spectral weight transfer likely associated with the electronic reconstruction by the vacancy sites, is systematically captured, as is the spatial extent of the vacancy region. Given the fact that a great variety of electronic materials will exhibit similarly smooth variation of the spectral responses due to random or engineered inhomogeneities in their structure, we believe our approach will be useful for systematic analysis of hyperspectral imaging with minimal prior knowledge, as well as prospective comparison of experimental measurements to theoretical calculations with explicit consideration of disorder.

**Introduction**

Electrical and mechanical measurements on the nanoscale underpin most widely used scanning probe microscopy methods as well as many future electronic devices. An outstanding challenge in such studies is systematic and quantitative interpretation of the measurements, often in the absence of well-defined analytical models. On the one hand, localizing the probed volume to near-atomic scale almost inevitably introduces unknowns into electronic, mechanical and chemical properties of the contacts, resulting in persistent and often time-varying uncertainty in the transfer function of the measurement. On the other hand, exponential amplification of physical properties in the measured observables is often the basis for enhanced sensitivity and spatial resolution achieved by imaging methods [1–3]. For example, exponential distance dependence of tunneling probability enabled scanning tunneling microscopy [4] and atomic resolution imaging of a broad variety structured and dynamics on surfaces [5–8]. The joint effects of uncertain and non-linear transfer function [9,10] generally complicate quantitative interpretation of data in scanning probe microscopy (STM) and related nanoscale device measurements. Moreover fully consistent theoretical treatment that accounts for the exact shape of the measurement probe, the non-equilibrium properties of electron tunneling and electronically excited states require extensive numerical simulation [9,10] that may be computationally too prohibitive to complement extensive experimentation. Therefore, a specific task of increasing importance is the analysis of hyperspectral data with minimum prior information [11]. The hyperspectral data volume can yield rich information even in the absence of a specific model, for example by identifying variability in the dataset or specific regions of interest.

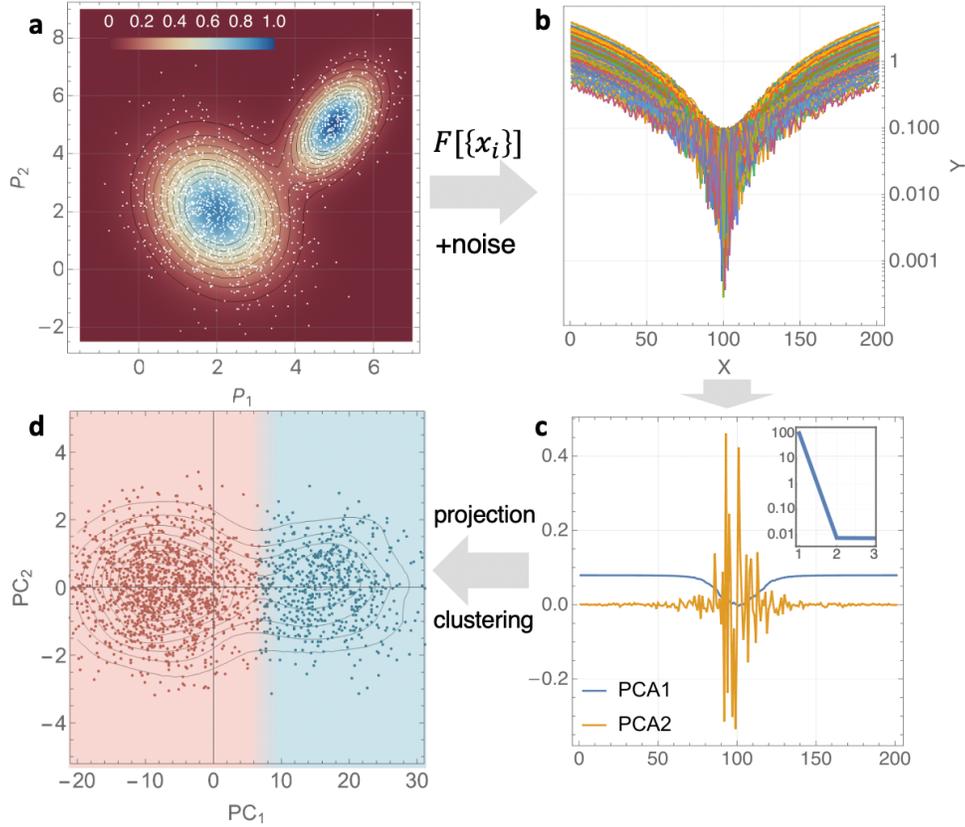

**Figure 1.** Conceptual goal of hyperspectral analysis of disorder is to infer the existence and structure of low dimensional latent space that effectively approximates the true parameter space, enabling reliable detection of disorder and further analysis. (a) Synthetic 2D parameter space, represented by multinormal distribution with two well separated centroids. (b) A family of polynomials parametrized by randomly sampled parameters from the synthetic space (a) (white dots in a). (c) First two principal components of the synthetic dataset (b), and the corresponding scree plot (inset), which show that projection on the first two principal components already capturers >99% of variability in the dataset. (d) Latent space of hyperspectral data projected onto first and second principal components (c), revealing structure very closely resembling the original parameter space. Two clusters (red and blue) are obtained by k-means clustering.

In the past several years, approaches utilizing dimensionality reduction and classification by clustering were successfully applied to hyperspectral data spatially resolved tunneling spectroscopy [12,13], force-distance spectroscopy [14,15], and switching spectroscopy of ferroelectrics [16,17]. The essential assumption of this approach is the existence of a low-dimensional parameter space from which measured spectra arise through convolution with the instrument transfer function. One common goal of such analysis is then to infer discrete representation of the hyperspectral data, for example in the form of several clusters of characteristic behaviors. Figure 1 illustrates this concept on synthetic data, wherein random samples {$P_i$} from the "true" 2D parameter space **Fig.1(a)** are used to construct a series of polynomial curves **Fig.1(b)**, representing the effect of the instrument function F[{$P_i$}] and added noise. Subsequently, principal component analysis projects the family of curves onto eigenvectors of the covariance matrix. The first two components already capture >99% of variability in the synthetic dataset (**Fig.1(c)**). The latent space of projections onto principal components **Fig.1(d)** can then be considered an effective representation of the data. Indeed, **Fig.1(d)** appears qualitatively very similar to **Fig.1(a)**, and can become nearly identical by affine transformation that rotates and stretches the parameter space. Applying clustering is also very effective in this case, drawing clear decision boundaries that partition the latent space into two parts, reminiscent of the two centers of the original distribution. Therefore, such analysis will work well if the true parameter space and therefore the derived latent space are both well separated. However, disorder in electronic materials can manifest in a broad range of behaviors, many of which will be perturbative rather than drastically changing [18–20]. In case of a sparse density of impurities, the number of data points originating from defect sites may be limited (it appears as an outlier), which may also confuse popular methods, such as k-means clustering [21]. And the validity of classification approaches can also be questioned, wherein partitioning the dataset into integer number of clusters inevitably introduces ambiguity near decision boundaries. As a result, good parameter separation in the latent space may be more of an exception for atomically-resolved microscopy data, and alternatives, such as probabilistic description or non-linear methods that can separate the latent space are needed for the likely more common case of smooth variation of properties across the dataset.

Here we demonstrate similarity analysis of hyperspectral data applied to detecting signatures of defects in tunneling spectroscopy of unconventional superconductors FeSe. Specifically, we identified the existence and electronic signature of iron defect sites in the hyperspectral cube, comprising spectra of energy-resolved density of states acquired over a spatial grid with near atomic-scale resolution. Despite apparent simplicity of the problem with just one kind of type of chemical vacancy, we show that conventional clustering methods may produce ambiguous if not conflicting results. We traced the origins of the ambiguity to the continuous and long-tailed structure of the distribution of Euclidian pairwise distances in the dataset. This conclusion was reinforced by the efficiency of ad-hoc rescaling of pairwise distances to identify the vacancy sites. Such analysis can be effective for low-defect scenarios even though it obscures well-defined geometric meaning of the latent space. To generalize, we applied effective weakly unsupervised neural network methodology, that explicitly learns pairwise similarities [22] and enables most effective analysis of our data-sets, even more so than many popular techniques for non-linear dimensionality reduction. We believe that similarity learning can play an important role in future analysis of a broad spectrum of problems in materials physics, such detecting and analyzing effects of disorder, as well as classical and quantum phase transitions.

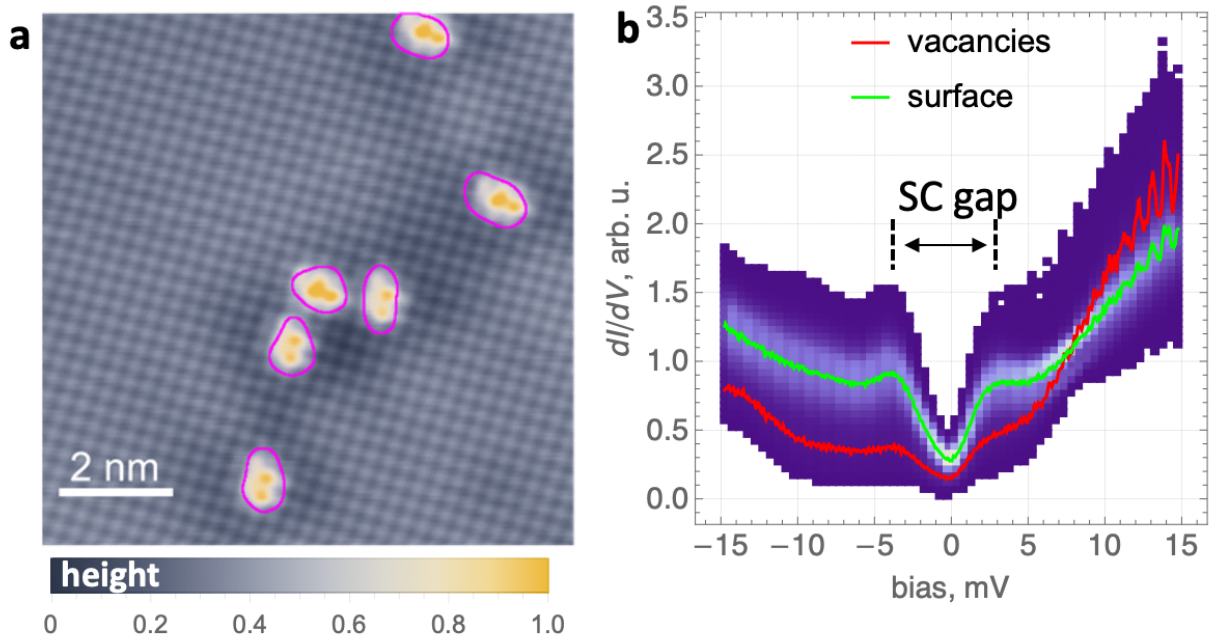

**Figure 2.** Tunneling microscopy and spectroscopy of as-cleaved FeSe surface, with a small number of Fe vacancies. (a) Constant-current STM image showing the atomic lattice of the top surface layer, and Fe vacancy sites which appear as clear double-protrusions (this is primarily an electronic effect); (b) tunneling spectroscopy near the Fermi level, revealing the ~ 4 meV wide superconducting gap, and a broad distribution of the tunneling conductance due to combined effects of measurement noise and vacancy sites. Individual spectra were acquired on a 100x100 pixel spatial grid, with 512 sampling points per energy interval from -15 mV to +15 mV. Purple in (b) is an energy-resolved histogram of all the spectra. Green and red mark the average spectra from the defect-free and defect regions of the surface, correspondingly. The vacancy sites were identified by segmentation analysis of an isoenergy slice of the hyperspectral volume at ~ -1 mV, and their approximate outlines are shown in purple in (a).

**Results and Discussion**

FeSe is an unconventional superconductor with a transition temperature of ~8K [23]. For this study, FeSe single crystals were grown out of KCl-AlCl$_3$ flux by liquid transport technique [24]. The surfaces were cleaved in ultrahigh vacuum at a temperature of ~150K prior to STM measurements. Owing to lack of dangling bonds, FeSe surfaces contain only a small number of well-defined defects [25–27]. Fe vacancies, for examples are readily detectable due their low surface density and the characteristic dumbbell shape (**Fig. 2a**), creating a model system for development of data analytics techniques. However, despite their chemical simplicity the vacancies in FeSe can have a profound impact on the electronic properties. For example, Fe vacancies locally modify magnetic ordering [28,29]. Meanwhile, Se vacancies have been predicted to generate effective hole doping through a combined effect of electron scattering, lattice strains and charge doping [30–32], contrary to intuitive expectation of electron-doping. The FeSe monolayer system can achieve almost 10x higher $T_c$ due to combined effect of doping and strain at the interface [33]. Meanwhile, the interest in FeSe/Te system has been vigorously renewed in lieu of predictions for the topologically protected superconducting state in proximity of 50/50 Se/Te ratio [34], where intrinsic disorder of solid solutions also emerges [35].

To characterize spectral signatures of the Fe vacancies, we acquired tunneling spectra on a grid of 100x100 points in a nearly pristine surface area with a few vacancies (**Fig. 2a**). Energy-resolved histogram of all the spectra clearly reveals the superconducting gap (**Fig. 2b**), as well as a rather broad distribution of density of states (about 4-fold variation), due to the effects of vacancies, electron scattering (resulting in Friedel oscillations), noise and other possible factors. Fe vacancies are already known to develop in-gap impurity states, and therefore increase the observed density of states inside the superconducting gap [36]. This effect enables partitioning of the spectra in the dataset into two groups – with and without vacancies – for example by segmentation of the iso-energy slice of the hyperspectral cube at an energy of ~ -1 mV. The corresponding outlines of the vacancy sites obtained by this segmentation are shown in **Fig. 2a** in purple. Quite notably, Fe vacancies significantly reconstruct the whole spectrum, not just the proximity of the superconducting gap (compare red and green curves in **Fig. 2b**).

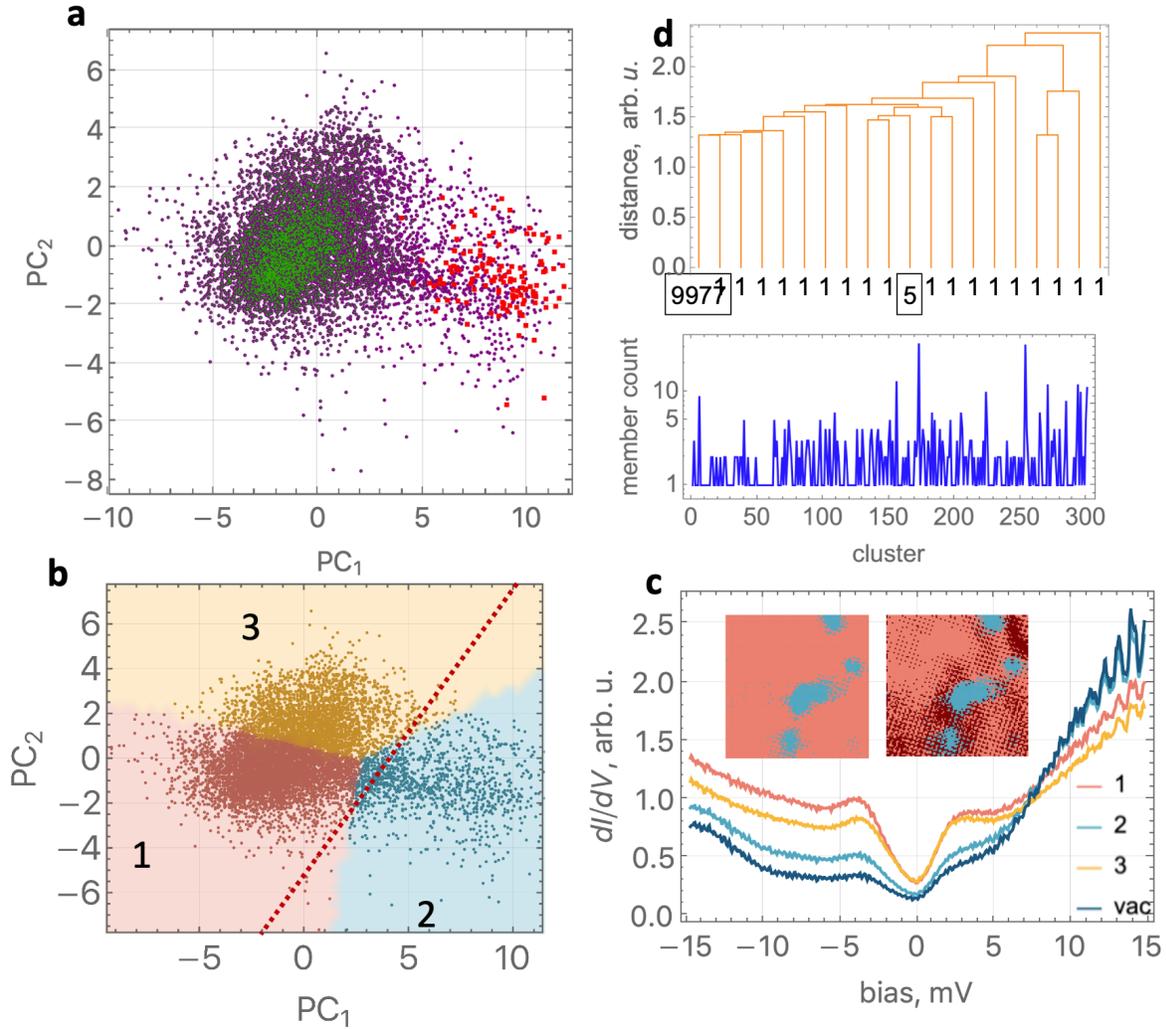

**Figure 3.** Dimensionality reduction and clustering applied to tunneling spectroscopy dataset in **Fig. 2**. (a) Principal component analysis (PCA): projection of the individual tunneling spectra into the latent space of two principal components ($PC_1, PC_2$) whose linear combination accounts for >90% of variability in the dataset. Green and red highlight projections due to surface and vacancy regions (determined by segmentation in **Fig. 2**), respectively. Purple are all the projections. (b) Partitioning of the PCA latent space by k-means clustering with 2 and 3 clusters. The decision boundary for 2 clusters is shown by the red dashed line. (c) Cluster centers (average IV curves within each cluster) for 3-means analysis in (b). Insets show spatial maps of clusters (left - 2, right - 3). The averaged I-V curve for the vacancy region (*vac* in the legend) from **Fig. 2** is shown for reference. (d) Results of agglomerative (hierarchical) clustering of the hyperspectral dataset, shown as a dendrogram. The y-axis shows linkage distance between individual clusters, while the numbers below show the number individual IV-curves in each shown branch of the dendrogram. Bottom panel shows the number of samples vs the hierarchical order of the specific cluster.

Although detecting vacancies in FeSe by visual inspection is remarkably easy, the multivariate workflow presented in **Fig. 1** fails to do so reliably in the latent space. This is the main focus of this article. Almost all methods presented below provide some level of separation in real-space. However, in a very general case – e.g. sparsely sampled dataset, disordered samples, multiple kinds of defects, real-space information may not be readily available. And we would like to find reliable approaches to separate characteristic behaviors in the spectral space alone. Eventually, both real-space and spectral space information should be utilized to maximize efficiency of analysis.

**Fig. 3a** shows projection of the hyperspectral dataset onto its first two principal components (that account for over 90% of the observed variance in the data, with 3 components accounting for over 99%). Locating the projections of the spectra due to vacancies (red) and pristine (green) surface sites shows that the two are separated in the latent space, albeit without clear break into distinct groups of points.

Application of two popular clustering methods to this latent space – k-means and agglomerative clustering - yields conflicting results: k-means points to few clusters in the dataset (**Fig. 3b**), while agglomerative clustering points to the opposite – dozens if not more distinct behaviors (**Fig. 3d**). K-means - a "top-down" algorithm - identifies k-number (where k is given) of centroids and iteratively refines their position until the net variances of the point-cluster distances are minimized [21,37]. Agglomerative (hierarchical) clustering builds a "bottom-up" cluster hierarchy [38], beginning with *n* clusters (where n is number of spectra in the dataset) and iteratively merging distinct clusters with the same linkage distance. At some intermediate value of linkage distance, a small number of clusters may be revealed. In our case, k-means is effective at approximate identification of the vacancy sites for 2 cluster partitioning, which is to be expected (see mask in the inset of **Fig. 4c**). Adding a 3$^{rd}$ cluster into partitioning, however, does not refine the location of the vacancy, instead partitioning the latent space region of the surface sites and assigns a few spurious locations to the vacancies (**Fig. 3c** inset). Meanwhile, hierarchical clustering, partitions the spectra into essentially arbitrarily many clusters, with dozens and even hundreds of clusters having approximately equal weighting. This is seen in **Fig. 3d**, where at no point up to even 300 clusters is the number of spectra in each cluster exceeds 10 spectra (and most often just one, **Fig. 3d** bottom panel). In an effective agglomerative clustering scenario, we would expect to have much more significant weighting for clusters already at the top of the hierarchy (e.g. for 3-4 cluster partitioning). In other words, according to agglomerative clustering, either there are hundreds of observable behaviors – which is unlikely given the origin of our data, or there are no significant clusters – i.e. vacancies and pristine surface are not statistically distinct – which contradicts direct observations.

Given the extensive amount of literature on clustering methods, such ambiguities are not uncommon [37]. It is rather clear that the Euclidian distance between individual spectra or between their projections in the latent space plays a key role in the subsequent performance of the clustering methods. In the synthetic dataset in **Fig. 1**, for example, there is a clear separation of the centroids in the parameter space, so that the clustering can readily succeed given enough samples from the distribution. No such clear separation is seen in **Fig. 3a**. Furthermore, we directly examined the distribution of Euclidian distances across the whole hyperspectral dataset (**Fig. 4a**). It is immediately apparent that the distribution is long tailed, which at least partially is caused by relatively small number of vacancy sites. The sparsity of the vacancies in the dataset creates an immediate problem for the k-means clustering method, which is most efficient when the number of samples from distinct behaviors is comparable in the data-set [21,37]. The long tail also confuses agglomerative clustering, essentially due too slow pace of agglomeration, so that inferred clusters end up containing very few spectra up to a very high level in the hierarchy (Fig. 3d).

Our reasoning can be further supported by an ad-hoc rescaling of the Euclidean distances between spectra. **Fig. 4b** plots the distributions of $log(d_{ij} + n)$ with n from 0.1 to 8 (where $d_{ij}$ is the Euclidean distance between a pair of spectra). In these cases, logarithmic rescaling symmetrizes the distance distribution by "contracting" the long tail of large relative distances while "expanding" smaller values. For *n* between 1 and 5, the distribution becomes not only much more symmetric, but it also clearly reveals a shoulder on the side of the main-peak (**Fig. 4b**. A weak shoulder is also visible in **Fig. 4a**). In this form, the modified distance distribution can be readily fit by two Gaussian distributions (**Fig. 4c**). The bimodality of the distribution then points to existence of at least two distinct groups of spectra, which are similar within the group (therefore yielding small relative Euclidean distance captured by the blue peak in **Fig. 4c**) and are dissimilar across the groups (with larger relative distance under the purple peak in **Fig. 4c**). Extending this logic, we can now take advantage of the sparsity of the vacancies on the surface to actually label the individual spectra as belonging to majority vs minority species. To this end, we first calculate distributions of log-Euclidean distance for each spectrum, and then effectively label each spectrum with its distance histogram. For majority (surface sites), the histograms are dominated by the first peak while for vacancies, the opposite is true (**Fig. 4d** inset). Visual inspection of the abundance maps of individual histograms in real space for the minority species in **Fig. 4d** indeed confirms our notions as well as the assignment of the assumed bimodal distribution in **Fig. 4c**.

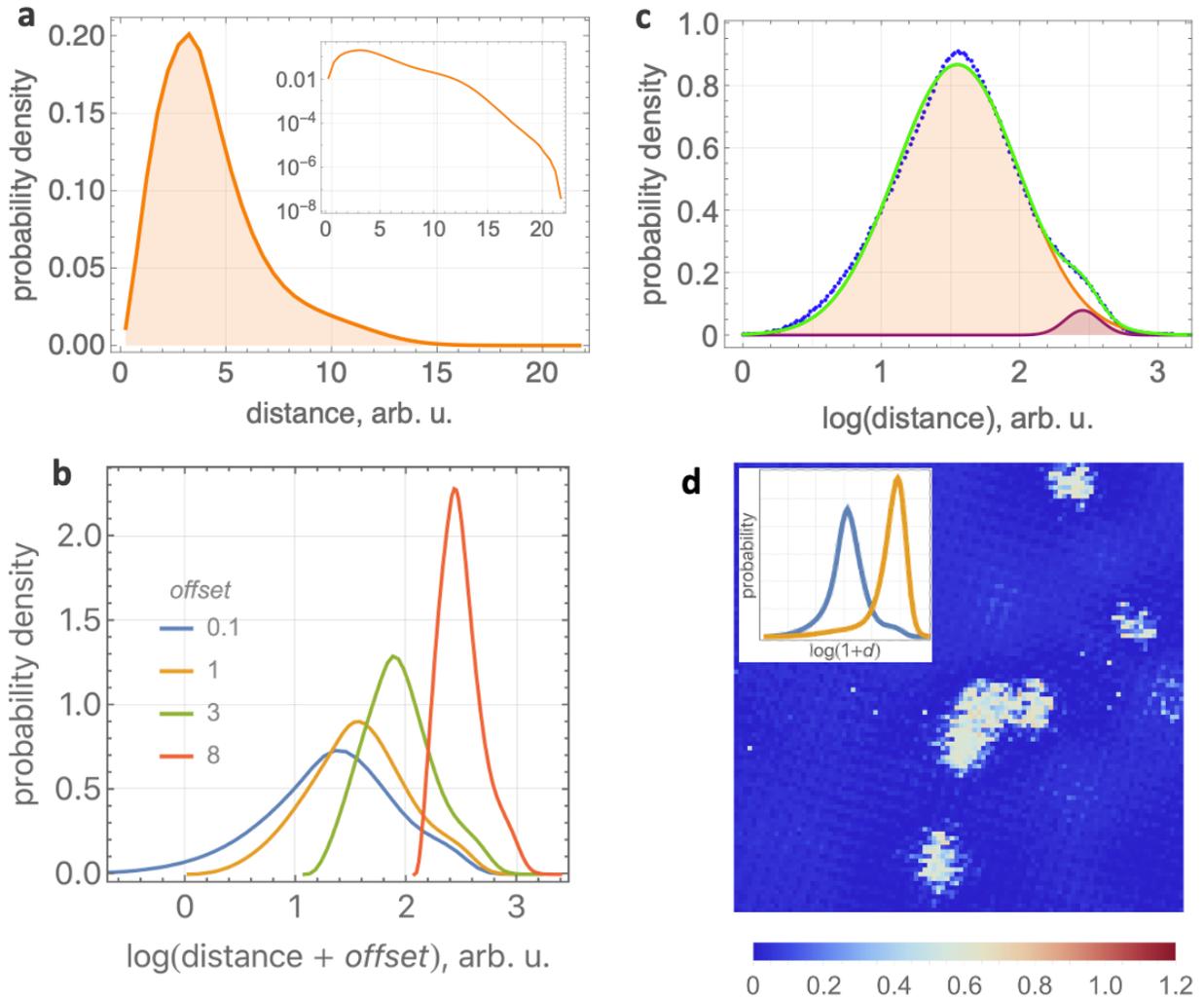

**Figure 4.** Statistics of pairwise Euclidean distances between single tunneling spectra in the FeSe dataset. (a) Distribution of Euclidean distances within the dataset from **Fig. 2b** exhibiting a long tail (inset shows distribution on the log scale). (b) Distributions of rescaled pairwise distances, with rescaling as $(\log(d_{ij} + offset)$. (c) Fitting rescaled distribution of the log-Euclidean distance by two normal distributions. (d) The abundance map of minority spectra in the dataset, obtained as value for each spectrum's distance histogram at rescaled distance value of $\log(d_{ij} + 1) = 2.5$. The inset clearly shows typical probability distributions for a majority (blue) and minority (orange) spectrum, with minority exhibiting a peak at approximately 2.5.

The rescaling described above, which amounts to non-linear geometric transformation of the data space, is effective, but it is also arbitrary and is likely not generalizable to more complex cases. We therefore sought non-linear dimensionality reduction methodologies, that can achieve non-linear and effective separation between tunneling spectra algorithmically. Non-linear dimensionality reduction encompasses a large range of algorithms, such as kernel PCA [39], TSNE (t-distributed stochastic neighbor embedding) [40], neural network autoencoders among others, each with its own approach to construction of the latent space.

To aid in evaluation of the effectiveness of a specific technique we tracked the relevant geometry of the latent space by plotting the Euclidean distance of each spectrum from the average in data-set spectrum vs similar metric in the latent space (Euclidean distances of coordinates of these spectra in a given latent space from the coordinates of the average spectrum). Examples of such a distance correlation plot, in the form of the density histogram, are shown in **Fig. 4d** for PCA and our logarithmic rescaling of the distances discussed above. PCA is a linear dimensionality reduction technique, and this is very well reflected by the distance correlation plot with a clear linear scaling of distances between original and reduced dimensionality representations (**Fig. 5a**). Deviation from linearity in this case is the effect of partial loss of information due to dimensionality reduction). Meanwhile, the logarithmic rescaling appears as a log-function, by design (**Fig. 5b**). The transition from the long-tailed to the more symmetric distribution of the relative distances is also apparent by following the redistribution of the point density upon log-rescaling.

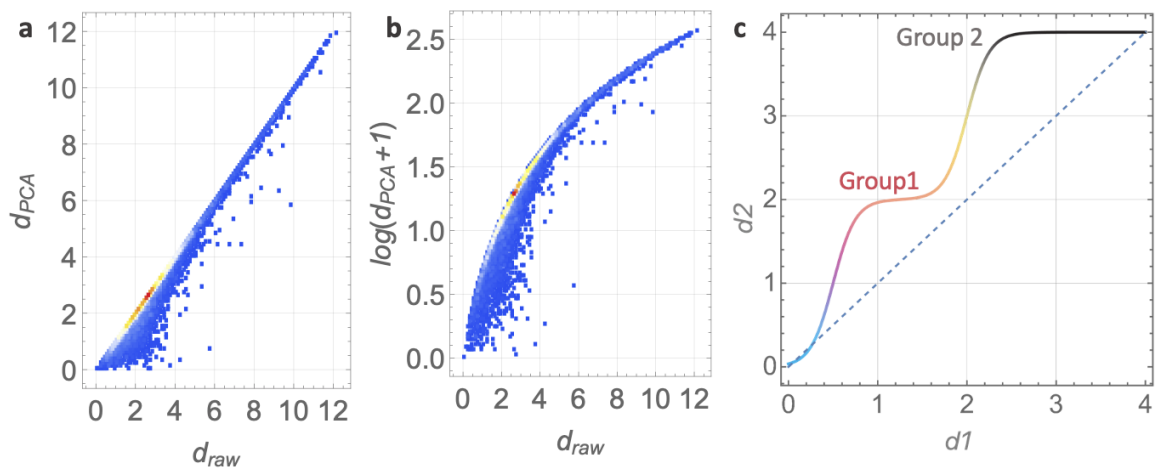

**Figure 5.** Distance correlation plot that compares distances between individual points within the data-set (or distance from the mean) among two data spaces. (a) Comparison of the original data

space before dimensionality reduction with the distance distribution in the latent space defined by first two principal components (**Fig. 3a**). (b) Similar comparison after logarithmic rescaling of the latent space. (c) Schematic for distance correlation plot for an idealized scenario of non-linear dimensionality reduction, where two (or more) groups of spectra would be clearly separated from each other by the appropriate distortion of the latent space.

Figure 6 shows the latent spaces and the distance correlation plots for the FeSe dataset generated by several popular methods for non-linear dimensionality reduction: kernel PCA [39,41], (**Fig. 6a**), non-negative matrix factorization (NMF) [42], U-MAP [43] (**Fig. 6c**) and deep autoencoder (**Fig. 6c**). All the methods introduce various non-linearities as seen from the correlation plots: kPCA "squeezes" the latent space as a whole (**Fig. 6a**), somewhat similar to the log-rescaling in Fig. 5. UMAP on the other hand, introduces non-linearity while overall maintaining the original distance scale (**Fig. 6c**). In neither of these cases is there a clear separation between separation between surface (blue) and vacancy (maroon) sites (**Fig. 6a-c**). One way such clear separation would be observed in the correlation plots is shown in **Fig. 5c**. The stepwise shape of the correlation plot would imply that non-linear dimensionality reduction partitions the data-set into several groups, and also separates them in the projected space. This specific shape is valid under the assumption that one group (group 2 in **Fig. 5c**) has a lot fewer data points than the other (e.g. group 1), which is typically the case for sparse density of vacancies and defects. Finally, the deep autoencoder does reconstruct the latent space, splitting it into approximately three regions, two of which (both surface sites, **Fig. 6d**) have comparable density. The vacancies however still maintain poorly separated. Thus, although non-linearities introduced by these methods have not yielded the sought-after result of clearly partitioning the data space, as is also evident from direct inspection of the latent spaces for all the methods (see insets). We do note that each of these methods has a set of hyper-parameters, and a pronounced sensitivity to specific choices of these parameters (e.g. compare kPCA results in **Fig. 6a** and **Fig. 6b**). However, we are seeking a more general solution. And given the relatively unsatisfactory performance of the unsupervised non-linear dimensionality reduction techniques, we next applied algorithms where some prior information about the properties of the spectra is passed to the algorithm.

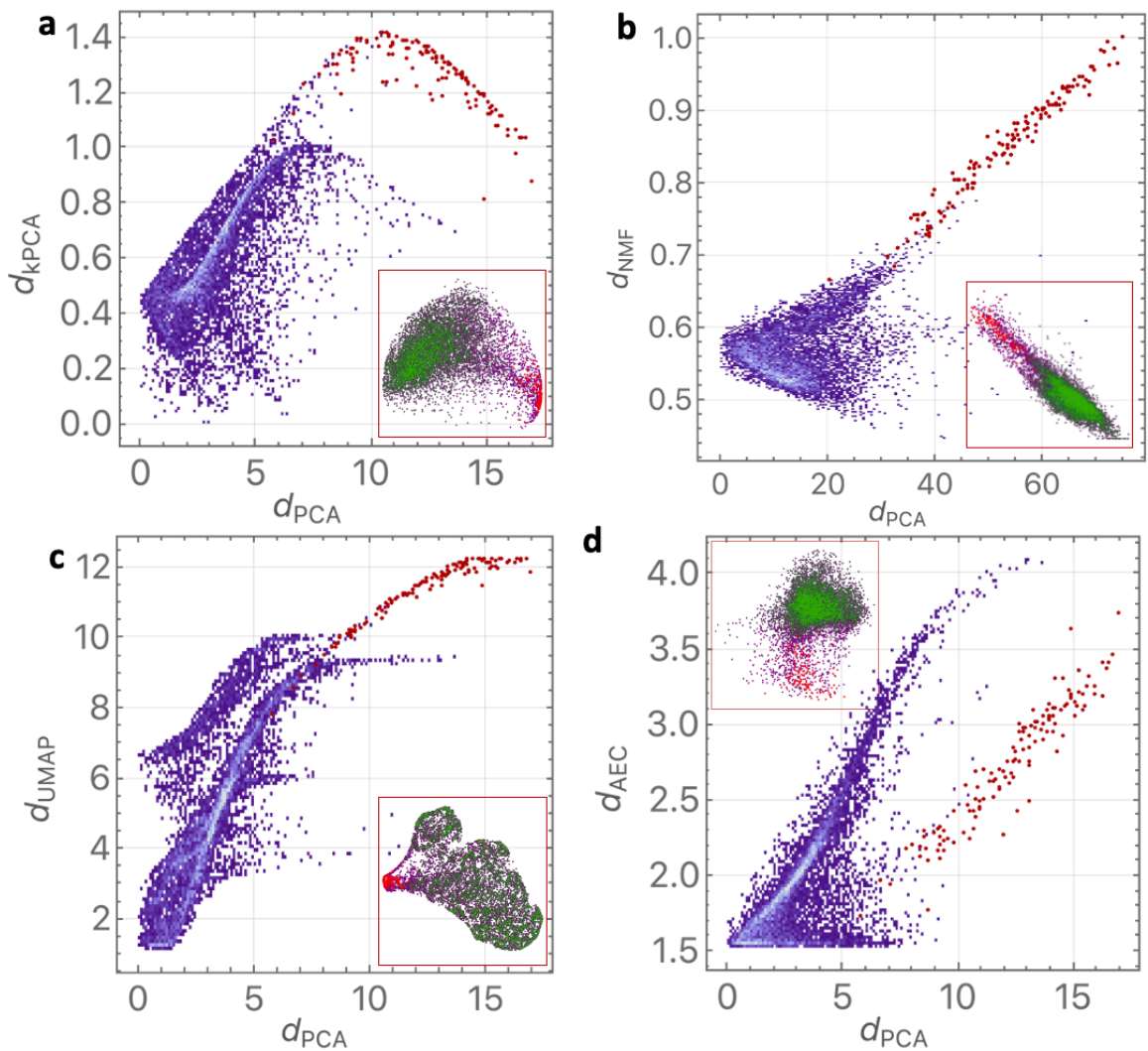

**Figure 6.** Distance correlation plots for kernel-PCA (a), non-negative matrix factorization (NMF, with implementation in scikit-learn [42] ), UMAP (c) and deep autoencoder (d) non-linear dimensionality reduction methods applied to the spectroscopy dataset. Blue and red accents in the main figure highlight the positions of the surface and vacancy sites correspondingly. The insets show 2D latent spaces obtained by each method. The parameters used for each plot are as follows: (a), (b), (c), (d).

We obtained satisfying performance with a twin neural network architecture (also referred to as Siamese networks), which was originally proposed by LeCun group [44] to achieve better classification of image data-sets. The schematic of such a network is shown in **Fig. 7a**. The network is being trained by propagating a pair of spectra through a pair of identical networks. The network pair has both identical topology and shared trainable weights. We chose a deep network consisting of three fully connected layers with dimensionality of 20→5→2 separated by non-linear activation functions. The network therefore maps the spectra into 2-dimensional projected space. Subsequently Euclidian distance is calculated between the outputs of the two networks (ED, Fig. 7a). The key to twin network performance is a contrastive loss (CL in **Fig. 7a**), which computes a loss based on the distance across the twin network and a target that specifies whether the distance should be minimized or maximized. The network can therefore be trained to create projected spaces where similar spectra are grouped together as much as possible, and distinct spectra are separated as much as possible – exactly the goal we desire for the tunneling spectra. Moreover, because the number of pairs scales as $N^2$ with the N-size of the data-set, twin networks work well with smaller size datasets characteristic of tunneling spectroscopy.

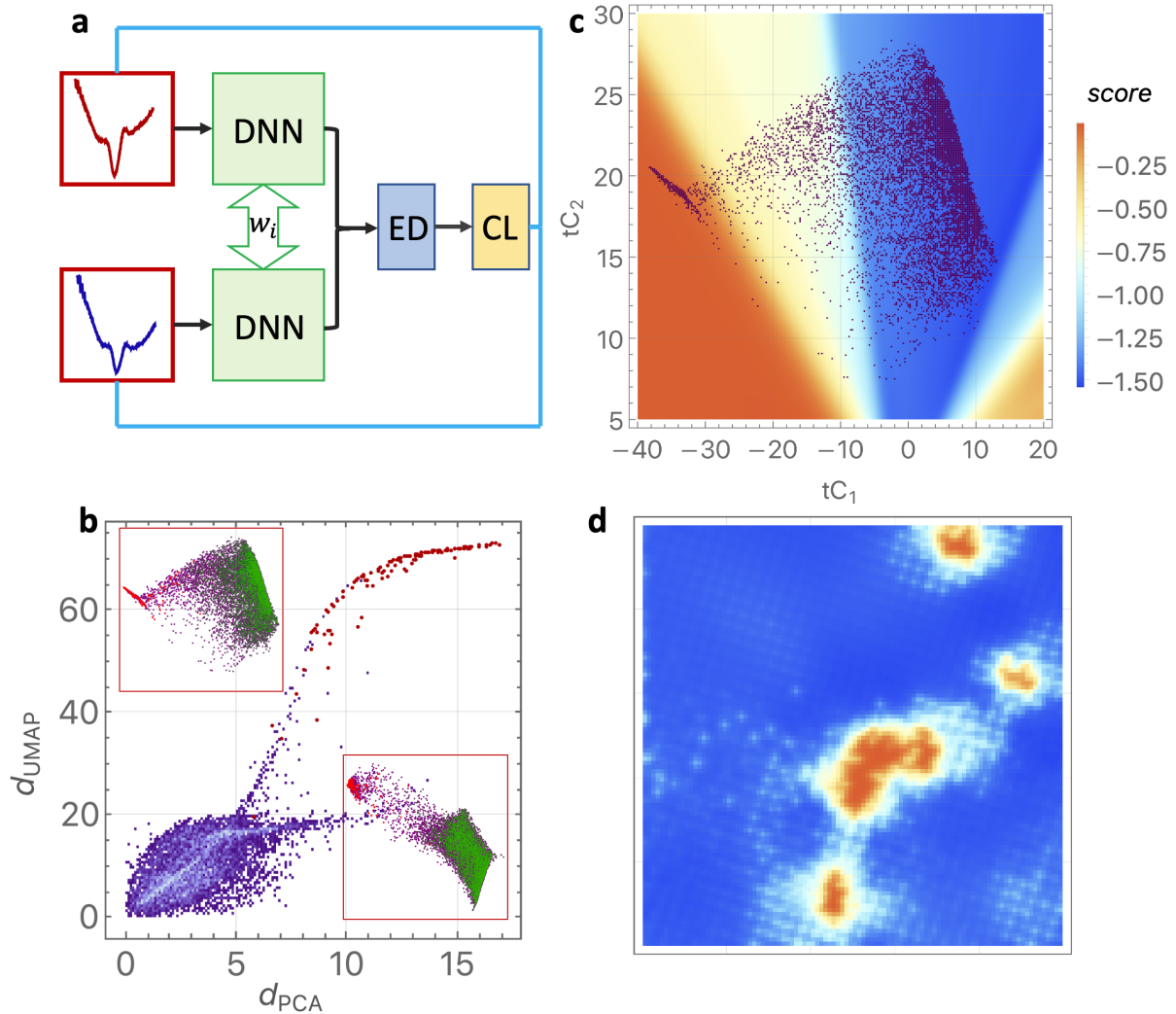

**Figure 7.** Application of similarity learning by twin (Siamese) network architecture for the tunneling spectroscopy dataset. (a) Schematic structure of the basic twin architecture (DNN – deep neural network, $w_i$ – shared weights, ED – Euclidean distance, CL -contrastive loss layer). (b) Distance correlation plot for a typical projection of the trained Siamese network applied to the dataset (insets show two different projections, obtained by slight variation of the parameters of the training set, such as the threshold Euclidean distance beyond which the spectra are considered to be distinct). (c) The result of applying two Siamese networks to the dataset in sequence. First network to project the data-set onto two components (tC1 and tC2) – the projection is the same as the distance correlation plot in Fig. 7b. 2$^{nd}$, simplified network, to create 1D scores for each of the projection, based on their proximity to each other in the projected space. The field of scores is superimposed on top of the projection as a colormap, while (d) shows the abundance map of the scores in real-space. Very clearly, the scores act as unique labels for individual spectra, clearly separating vacancy, surface and intermediate sites. The image has been slightly smoothed by Gaussian convolution for presentation purposes.

The unsupervised input into the twin network is a binary choice of whether the two spectra are of the same overall group. In our case we simply chose a particular threshold for the Euclidean distance in the original hyperspectral dataset, to partition the pairs within the dataset. As seen **Fig. 7b**, vacancies (red) become well separated from surface sites in the projected space of the twin network (green). This is also clearly seen in the distance correlation, which bears closest similarity to idealized scenario in **Fig. 5c**. The training is also sensitive to hyper-parameters, although producing qualitatively similar results. **Fig. 7b** shows two qualitatively similar projected spaces obtained in different initialization and training runs (with randomly generated training sets in each case).

The twin architecture then enables categorization of the projected and latent spaces, as shown in **Fig. 7c**, wherein another network with smaller and even simpler (10→5→1) topology was applied to the projected space in the upper corner of **Fig. 7b**. The network in this case outputs just one number, providing a continuously varying score for each spectrum based on its proximity to other spectra in the projected space. As seen in **Fig. 7c**, twin networks very effectively categorize the latent space, drawing clear decision boundaries for the surface (blue) and vacancy (red) sites, and rather remarkably outlining the transition region between the two (yellow). Because this type of classification avoids discretization of the clustering procedure, the surface map of the scores both confirms its assignment and separates the vacancy, surface sites, and the transition region in both projected and real spaces.

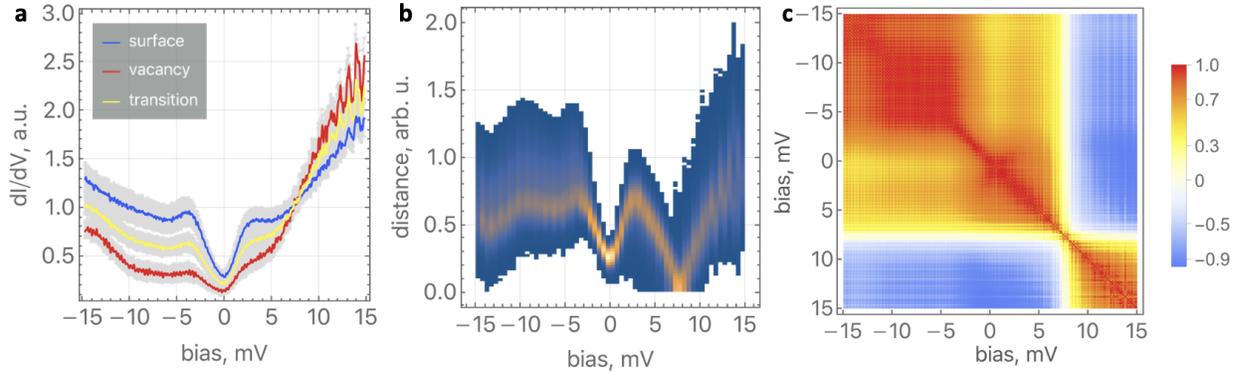

**Figure 8.** Spectral weight transfer due to Fe vacancies in FeSe as seen by tunneling spectroscopy. (a) Average spectra due to surface, vacancy and transition regions (red, blue and yellow) identified by the appropriate scores in **Fig. 7d**. (b) Energy-resolved Euclidean distances between the collection of surface and vacancy spectra, plotted as a density histogram (yellow corresponds to high abundance). (c) Correlation plot for all spectra in the data-set, revealing two distinct anticorrelated energy windows for spectral changes (from -15 mV to -5 mV and from 7 mV to 15 mV), as well as window of intermediate correlation (between about 0 mV and 7 mV). This plot statistically confirms that Fe vacancies reduce the density of filled states and increase the density of empty states compared to pristine surface.

The key prospective advantage of being able to identify distinct tunneling spectra as well as transitions between them is analysis of the spectral weight transfer associated with disorder in any doped, but particularly highly correlated electronic materials [19,20,45]. **Fig. 8** shows the tunneling spectra of the vacancies, surface sites and transition region averaged within their characteristic score values as determined in **Fig. 7c**. For added clarity, we also show the Euclidean distances between the spectra in **Fig. 8b** and a correlation plot for the whole dataset in **Fig. 8c**. The spectral intensity is redistributed in the whole probed energy window +/-15 meV, not just around the superconducting gap. The correlation plot reveals that the changes from -15 mV to ~ +7 mV are anticorrelated with those from +7 mV to 15 mV, wherein suppression of the density of states in one region causes enhancement in the other. The redistribution of the spectral weight may signal significant electronic reconstruction associated with the vacancy site, that goes beyond charge doping expected of impurities and intentional dopants. In fact, Se vacancies were predicted to cause significant reconstruction within as much as +/- 2 eV window around Fermi level, causing substantially weaker doping of the Fe d-bands than expected based on stochiometric arguments [31]. The detailed understanding and theoretical analysis of this phenomenon for the Fe vacancies is left for future work. However, we do note that the analysis presented here may also conceptually translate to analysis of the first principles data, providing a path to compare theory and experiment through the effects of disorder on the electronic structure.

**Conclusions**

To conclude, we applied several dimensionality reduction techniques toward effective measurement of effects of disorder on the electronic properties of FeSe superconductor using tunneling spectroscopy. It is shown that a combination of continuous variation of the electronic density of states, expected for atomic and nanoscale defects and impurities, combined with relatively low defect density contribute to inconsistent performance of discretizing clustering approaches to the structure of the hyperspectral dataset as well as identification of the vacancy sites. Meanwhile, several common nonlinear dimensionality reduction algorithms were not effective in separation of the vacancy sites in the latent space, that could potentially improve the follow-on clustering. Instead, twin (Siamese) neural network architecture, that learns the similarity directly from the data and amplifies the similarities and dissimilarities during training was shown to be effective at both creating well-separated projected space of the tunneling spectroscopy dataset, as well as subsequently categorizing tunneling with continuous single-valued scores. The immediate advantage of this approach is the ability not only to detect and describe the properties of defects and impurities, but also those of the transition region between distinct behaviors in the dataset. This is arguably a key advantage over discretization of the hyperspectral dataset with clustering methods, wherein neither the uniform state within a cluster nor abrupt cluster boundaries are physically sound. Moreover, twin networks are well adapted to work with smaller size datasets characteristic of tunneling spectroscopy, due to $N^2$ scaling of the training set size with the N-size of the dataset. In the specific case of FeSe, we have found that Fe impurities modify the density of states in a broad window around Fermi level, far beyond the superconducting gap. This may be an indication of an even more dramatic electronic reconstruction, qualitatively similar to earlier predictions of the effect of Se vacancies. The flexibility of twin network architecture not only enables capturing such effects in a more rigorous and systematic approach, but prospectively further comparison of theoretical and experimental methodologies for analysis of disorder on similar grounds. Ultimately, these methods will serve to better understand the effects of disorder in complex electronic materials, toward tailored fundamental and applied properties for advancement of quantum electronics.

**Methods**

Tunneling spectroscopy and microscopy was carried out using Specs Joule-Thomson STM microscope, operating at 4.5K base temperature. Freshly cleaved surfaces were obtained by delaminating a small sample of FeSe in low-vacuum of ~ $10^{-6}$ Torr, followed by rapid sample transfer to the cryogenic chamber. Data analysis presented above was carried out using primarily Wolfram Language, with additional use of Python libraries for non-linear dimensionality reduction. All the data and codes are available upon reasonable request.

**Acknowledgements**

We thank Rama Vasudevan for discussion and improvement of the manuscript. Research sponsored by Division of Materials Science and Engineering, Basic Energy Sciences, Office of Science, US Department of Energy. Experiments were carried out as part of a user project at the Center for Nanophase Materials Sciences, Oak Ridge National Laboratory. This research used resources of the Compute and Data Environment for Science (CADES) at the Oak Ridge National Laboratory, which is supported by the Office of Science of the U.S. Department of Energy under Contract No. DE-AC05-00OR22725.

**Author contribution**

PM guided research, helped to acquire data, carried out machine learning and interpretation. JY and BS grew FeSe crystals. JW acquired the data and carried out initial data analysis. All authors contributed to writing of the manuscript.

**References**

[1] F. J. Giessibl, *Advances in Atomic Force Microscopy*, Rev. Mod. Phys. **75**, 949 (2003).

[2] L. Olesen, M. Brandbyge, M. R. Sørensen, K. W. Jacobsen, E. Lægsgaard, I. Stensgaard, and F. Besenbacher, *Apparent Barrier Height in Scanning Tunneling Microscopy Revisited*, Physical Review Letters **76**, 1485 (1996).

[3] P. Maksymovych, M. Pan, P. Yu, R. Ramesh, A. P. Baddorf, and S. V. Kalinin, *Scaling and Disorder Analysis of Local I-V Curves from Ferroelectric Thin Films of Lead Zirconate Titanate*, Nanotechnology **22**, 254031 (2011).

[4] G. Binnig and H. Rohrer, *Scanning Tunneling Microscopy---from Birth to Adolescence*, Rev. Mod. Phys. **59**, 615 (1987).

[5] L. Grill, *Functionalized Molecules Studied by STM: Motion, Switching and Reactivity*, J. Phys.: Condens. Matter **20**, 053001 (2008).

[6] P. Maksymovych, *Excitation and Mechanisms of Single Molecule Reactions in Scanning Tunneling Microscopy*, in *Scanning Probe Microscopy of Functional Materials*, edited by S. V. Kalinin and A. Gruverman (Springer New York, 2010), pp. 3–37.

[7] Ø. Fischer, M. Kugler, I. Maggio-Aprile, C. Berthod, and C. Renner, *Scanning Tunneling Spectroscopy of High-Temperature Superconductors*, Rev. Mod. Phys. **79**, 353 (2007).

[8] J. E. Hoffman, *Spectroscopic Scanning Tunneling Microscopy Insights into Fe-Based Superconductors*, Reports on Progress in Physics **74**, 124513 (2011).

[9] H. Lin, J. M. C. Rauba, K. S. Thygesen, K. W. Jacobsen, M. Y. Simmons, and W. A. Hofer, *First-Principles Modelling of Scanning Tunneling Microscopy Using Non-Equilibrium Green's Functions*, Front. Phys. China **5**, 369 (2010).

[10] J. M. Blanco, F. Flores, and R. Pérez, *STM-Theory: Image Potential, Chemistry and Surface Relaxation*, Progress in Surface Science **81**, 403 (2006).

[11] S. V. Kalinin, E. Strelcov, A. Belianinov, S. Somnath, R. K. Vasudevan, E. J. Lingerfelt, R. K. Archibald, C. Chen, R. Proksch, N. Laanait, and S. Jesse, *Big, Deep, and Smart Data in Scanning Probe Microscopy*, ACS Nano **10**, 9068 (2016).

[12] M. Ziatdinov, A. Maksov, L. Li, A. S. Sefat, P. Maksymovych, and S. V. Kalinin, *Deep Data Mining in a Real Space: Separation of Intertwined Electronic Responses in a Lightly Doped BaFe 2 As 2*, Nanotechnology **27**, 475706 (2016).

[13] A. Belianinov, P. Ganesh, W. Lin, B. C. Sales, A. S. Sefat, S. Jesse, M. Pan, and S. V. Kalinin, *Research Update: Spatially Resolved Mapping of Electronic Structure on Atomic Level by Multivariate Statistical Analysis*, APL Materials **2**, 120701 (2014).

[14] B. J. Albers, T. C. Schwendemann, M. Z. Baykara, N. Pilet, M. Liebmann, E. I. Altman, and U. D. Schwarz, *Three-Dimensional Imaging of Short-Range Chemical Forces with Picometre Resolution*, Nature Nanotechnology **4**, 5 (2009).

[15] E. Minelli, G. Ciasca, T. E. Sassun, M. Antonelli, V. Palmieri, M. Papi, G. Maulucci, A. Santoro, F. Giangaspero, R. Delfini, G. Campi, and M. De Spirito, *A Fully-Automated Neural Network Analysis of AFM Force-Distance Curves for Cancer Tissue Diagnosis*, Appl. Phys. Lett. **111**, 143701 (2017).

[16] J. C. Agar, Y. Cao, B. Naul, S. Pandya, S. van der Walt, A. I. Luo, J. T. Maher, N. Balke, S. Jesse, S. V. Kalinin, R. K. Vasudevan, and L. W. Martin, *Machine Detection of Enhanced Electromechanical Energy Conversion in PbZr0.2Ti0.8O3 Thin Films*, Advanced Materials **30**, 1800701 (2018).

[17] S. M. Neumayer, S. Jesse, G. Velarde, A. L. Kholkin, I. Kravchenko, L. W. Martin, N. Balke, and P. Maksymovych, *To Switch or Not to Switch - a Machine Learning Approach for Ferroelectricity*, Nanoscale Advances **2**, 2063 (2020).

[18] F. Ming, S. Johnston, D. Mulugeta, T. S. Smith, P. Vilmercati, G. Lee, T. A. Maier, P. C. Snijders, and H. H. Weitering, *Realization of a Hole-Doped Mott Insulator on a Triangular Silicon Lattice*, Phys. Rev. Lett. **119**, 266802 (2017).

[19] B. W. Hoogenboom, C. Berthod, M. Peter, Ø. Fischer, and A. A. Kordyuk, *Modeling Scanning Tunneling Spectra of Bi 2 Sr 2 CaCu 2 O 8 + δ*, Phys. Rev. B **67**, 224502 (2003).

[20] M. J. Rozenberg, G. Kotliar, and H. Kajueter, *Transfer of Spectral Weight in Spectroscopies of Correlated Electron Systems*, Phys. Rev. B **54**, 8452 (1996).

[21] A. K. Jain, *Data Clustering: 50 Years beyond K-Means*, Pattern Recognition Letters **31**, 651 (2010).

[22] S. Chopra, R. Hadsell, and Y. LeCun, *Learning a Similarity Metric Discriminatively, with Application to Face Verification*, in *2005 IEEE Computer Society Conference on Computer Vision and Pattern Recognition (CVPR'05)*, Vol. 1 (2005), pp. 539–546 vol. 1.

[23] A. Kreisel, P. J. Hirschfeld, and B. M. Andersen, *On the Remarkable Superconductivity of FeSe and Its Close Cousins*, Symmetry **12**, 9 (2020).

[24] J.-Q. Yan, B. C. Sales, M. A. Susner, and M. A. McGuire, *Flux Growth in a Horizontal Configuration: An Analog to Vapor Transport Growth*, Phys. Rev. Materials **1**, 023402 (2017).

[25] T. Berlijn, H.-P. Cheng, P. J. Hirschfeld, and W. Ku, *Do Se Vacancies Electron Dope Monolayer FeSe?*, Physical Review B **89**, (2014).

[26] X. Liu, L. Zhao, S. He, J. He, D. Liu, D. Mou, B. Shen, Y. Hu, J. Huang, and X. J. Zhou, *Electronic Structure and Superconductivity of FeSe-Related Superconductors*, J. Phys.: Condens. Matter **27**, 183201 (2015).

[27] D. Huang, T. A. Webb, C.-L. Song, C.-Z. Chang, J. S. Moodera, E. Kaxiras, and J. E. Hoffman, *Dumbbell Defects in FeSe Films: A Scanning Tunneling Microscopy and First-Principles Investigation*, Nano Letters **16**, 4224 (2016).

[28] S. Qiao, P. Zhang, H. Ding, S. Zhang, L. Liang, Z. Zhang, X. Long, X. Chen, J. Lu, and J. Wu, *Fingerprint of Checkerboard Antiferromagnetic Order in FeSe Monolayer Due to Magnetic-Electric Correlation*, Materials Today **41**, 44 (2020).

[29] W. Li, Y. Zhang, P. Deng, Z. Xu, S.-K. Mo, M. Yi, H. Ding, M. Hashimoto, R. G. Moore, D.-H. Lu, X. Chen, Z.-X. Shen, and Q.-K. Xue, *Stripes Developed at the Strong Limit of Nematicity in FeSe Film*, Nature Phys **13**, 957 (2017).

[30] T. Berlijn, P. J. Hirschfeld, and W. Ku, *Effective Doping and Suppression of Fermi Surface Reconstruction via Fe Vacancy Disorder in ${\mathbf{K}}_{x}{\mathrm{Fe}}_{2\mathbf{\ensuremath{-}}y}{\mathrm{Se}}_{2}$*, Phys. Rev. Lett. **109**, 147003 (2012).

[31] T. Berlijn, H.-P. Cheng, P. J. Hirschfeld, and W. Ku, *Doping Effects of Se Vacancies in Monolayer FeSe*, Physical Review B **89**, (2014).

[32] K.-W. Lee, V. Pardo, and W. E. Pickett, *Magnetism Driven by Anion Vacancies in Superconducting $\ensuremath{\alpha}\text{-FeSe}_{1\ensuremath{-}x}$*, Phys. Rev. B **78**, 174502 (2008).

[33] D. Huang and J. E. Hoffman, *Monolayer FeSe on SrTiO3*, Annu. Rev. Condens. Matter Phys. **8**, 311 (2017).

[34] Z. Wang, P. Zhang, G. Xu, L. K. Zeng, H. Miao, X. Xu, T. Qian, H. Weng, P. Richard, A. V. Fedorov, H. Ding, X. Dai, and Z. Fang, *Topological Nature of the ${\mathrm{FeSe}}_{0.5}{\mathrm{Te}}_{0.5}$ Superconductor*, Phys. Rev. B **92**, 115119 (2015).

[35]  L. Vlcek, A. Maksov, M. Pan, R. K. Vasudevan, and S. V. Kalinin, *Knowledge Extraction from Atomically Resolved Images*, ACS Nano **11**, 10313 (2017).
[36]  L. Jiao, S. Rößler, C. Koz, U. Schwarz, D. Kasinathan, U. K. Rößler, and S. Wirth, *Impurity-Induced Bound States inside the Superconducting Gap of FeSe*, Phys. Rev. B **96**, 094504 (2017).
[37]  Y. P. Raykov, A. Boukouvalas, F. Baig, and M. A. Little, *What to Do When K-Means Clustering Fails: A Simple yet Principled Alternative Algorithm*, PLoS ONE **11**, e0162259 (2016).
[38]  F. Murtagh and P. Legendre, *Ward's Hierarchical Agglomerative Clustering Method: Which Algorithms Implement Ward's Criterion?*, J Classif **31**, 274 (2014).
[39]  H. Hoffmann, *Kernel PCA for Novelty Detection*, Pattern Recognition **40**, 863 (2007).
[40]  L. van der Maaten and G. Hinton, *Visualizing Data Using T-SNE*, Journal of Machine Learning Research **9**, 2579 (2008).
[41]  B. Scholkopf, A. Smola, and K.-R. Müller, *Kernel Principal Component Analysis*, in *Advances in Kernel Methods - Support Vector Learning* (MIT Press, 1999), pp. 327–352.
[42]  F. Pedregosa, G. Varoquaux, A. Gramfort, V. Michel, B. Thirion, O. Grisel, M. Blondel, P. Prettenhofer, R. Weiss, V. Dubourg, J. Vanderplas, A. Passos, D. Cournapeau, M. Brucher, M. Perrot, and E. Duchesnay, *Scikit-Learn: Machine Learning in Python*, Journal of Machine Learning Research **12**, 2825 (2011).
[43]  L. McInnes, J. Healy, and J. Melville, *UMAP: Uniform Manifold Approximation and Projection for Dimension Reduction*, ArXiv:1802.03426 [Cs, Stat] (2020).
[44]  G. Koch, R. Zemel, and R. Salakhutdinov, *Siamese Neural Networks for One-Shot Image Recognition*, 8 (n.d.).
[45]  T. C. Koethe, Z. Hu, M. W. Haverkort, C. Schüßler-Langeheine, F. Venturini, N. B. Brookes, O. Tjernberg, W. Reichelt, H. H. Hsieh, H.-J. Lin, C. T. Chen, and L. H. Tjeng, *Transfer of Spectral Weight and Symmetry across the Metal-Insulator Transition in $VO_2$*, Phys. Rev. Lett. **97**, 116402 (2006).